\documentclass[conference]{IEEEtran}
\IEEEoverridecommandlockouts
\usepackage{cite}
\usepackage{amsmath,amssymb,amsfonts}
\usepackage{algorithmic}
\usepackage{graphicx}
\usepackage{textcomp}
\usepackage{xcolor}
\usepackage{booktabs}
\usepackage{setspace}
\usepackage{float}
\usepackage[hidelinks]{hyperref}
\def\BibTeX{{\rm B\kern-.05em{\sc i\kern-.025em b}\kern-.08em
    T\kern-.1667em\lower.7ex\hbox{E}\kern-.125emX}}
\begin{document}

\title{\vspace{20pt}Memory Optimization for Convex Hull Support Point Queries\\}

\author{\IEEEauthorblockN{Michael Greer}
\IEEEauthorblockA{michael@kinematiclaboratories.com \\
\textit{Kinematic Labs} \\
Boulder, Colorado \\
}
}

\maketitle

\begin{abstract}
\textit{This work has been submitted to the IEEE for possible publication. Copyright may be transferred without notice, after which this version may no longer be accessible.}
Support point queries are a critical part of many collision detection pipelines, including those for robotics and real-time graphical applications. This paper proposes several memory layout optimizations to speed up support point queries on convex hulls. These methods are implemented and tested on a variety of different hardware models, with a decrease in processing time of up to five times compared to current approaches. The results in this paper can be integrated with existing physics libraries with minimal effort.
\end{abstract}


\section{Introduction}

Interest in real-time robotic path planning is increasing as robotic systems become more ubiquitous and flexible, and with this advent comes the need for computationally efficient real-time collision modeling. For the past several decades, the default method has been to model rigid body simulations as interactions between one or more convex shapes, due to the efficiency of the GJK algorithm \cite{gilbert_fast_1987}. In recent years, much effort has gone into refining this algorithm: these efforts have targeted both the simplex distance sub-problem \cite{montanari_improving_nodate} as well as guided approaches for the GJK algorithm at a higher level \cite{montaut_gjk_2024} \cite{montaut_collision_2022}. However, fewer efforts have considered the third necessary component of the GJK algorithm: the calculation of the support point function.

One of the most desirable qualities of the GJK algorithm is its generality. The algorithm can process distance and intersection queries between any two convex shapes, so long as a support point function can be defined on each shape. The support point function accepts a search direction and computes the point on the shape that is the farthest along that direction in space. Alternatively, one may think of the search vector, instead, as the supporting plane perpendicular to this vector. In this case, the support point is the point on the shape where the supporting plane is tangent, or the point such that the plane through that point perpendicular to the search vector only intersects the shape at that supporting point. The support point function is trivially defined and closed-form for simple shapes such as spheres, capsules, and cones, but the function has also been defined for more complex shapes such as signed distance functions \cite{lopez-adeva_fernandez-layos_convex_2024} and NURBS surfaces \cite{dyllong_distance_nodate}.

One type of shape that has seen frequent use is the convex polytope, or convex hull. This representation is very flexible, as it can arbitrarily approximate any of the other aforementioned shapes. Additionally, convex hulls (and collections of convex hulls computed as a convex decomposition \cite{lien_approximate_2007}) can be evaluated directly from triangle meshes and point clouds, both of which are frequently utilized in robotic path planning contexts. However, this comes with two significant drawbacks: greatly increased storage size in memory and higher computation time, since the size of the data structure and resulting computation is generally proportional to the number of vertices on the shape. Previous work has sought to optimize these convex hulls by reducing the number of vertices, while minimizing the deviation from the original shape.

This paper demonstrates ways to expedite the computation of convex hull support points on scalar and superscalar processing devices (i.e. traditional CPUs) by improving the cache behavior of the underlying data structures. I propose three ways to improve the memory layout of the convex hull data: linking vertices with artificial internal edges, traversing the face normals of the convex hull facets to exploit consistent neighbor count, and compressing these normals using a reduced spherical representation to represent more faces per cache line. I implement this and other approaches and compare them to existing methods. On average, the approaches proposed here provide a reduction in processing time of 1.5 to 5 times compared to approaches used in popular physics simulation libraries such as Bullet.

\section{Background}

\subsection{Hill Climbing}

To measure a point along a vector, the point is projected onto the infinite line defined by the vector; this is identical to taking the dot product between the vector and the point, which provides the distance along the vector where the point lies in space. Consider a collection of points in space with no underlying relationship or connectivity information. It is possible to compute the support point of such a collection, but it would require computation of the dot product between every point and the search query. Na\text{\"i}ve support point functions on convex hulls behave in exactly this way: by computing the dot product between every vertex and the search vector.

\begin{figure}
    \centering
    \includegraphics[width=0.5\linewidth]{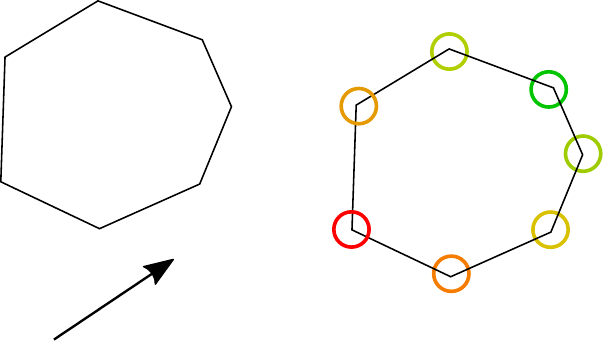}
    \caption{Convex Hull with Every Dot Product Evaluated}
    \label{fig:every-dot-product}
\end{figure}

There already exists an improvement on this approach that exploits the connectivity data of the convex hull: hill climbing \cite{sato_efficient_1996}. The optimization space of the vertices measured against a support vector is convex. This means that repeatedly moving in a locally optimal search direction is guaranteed to result in a final global optimum. This locally optimal step is implemented using the edge data of the convex hull. The algorithm compares each point to its neighbors. If the dot product of the current point with the search vector is greater than the dot product of all of it's neighbors, then it must be the corresponding support point, if not, the algorithm selects the neighbor with the greatest dot product, and the process repeats.

\begin{figure}
    \centering
    \includegraphics[width=0.8\linewidth]{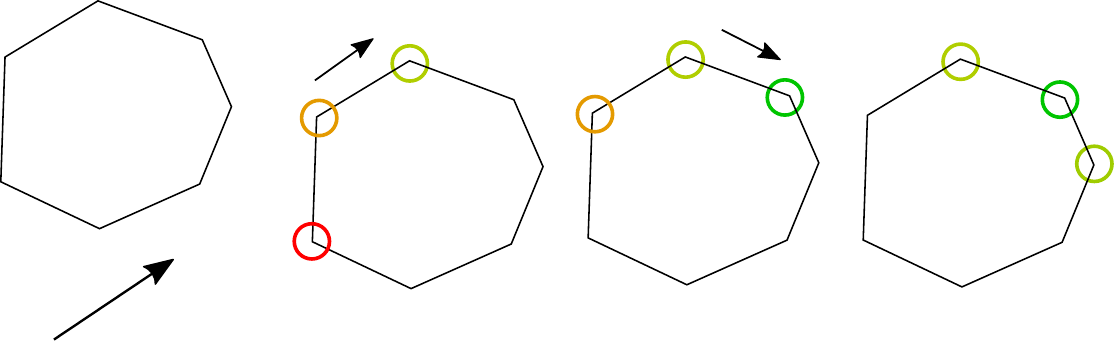}
    \caption{A Sequence of Hill Climbing Steps}
    \label{fig:hill-climbing}
\end{figure}

These two strategies are shown in Figures \ref{fig:every-dot-product} and \ref{fig:hill-climbing}, with red representing a low dot product value and green representing a high dot product value. To implement the hill climbing algorithm, any data structure that describes a convex hull must not only contain all of the cartesian coordinates of the vertices, but also all of the neighbors of each point. This is the absolute minimum information required for this algorithm to operate.

\subsection{Precomputation}

Another common way to reduce computation time is to warm-start the algorithm using precomputed start vertices. These vertices provide stronger initial starting points than simply choosing a random point on the shape. Each shape in this implementation has only six warmstart points aligned with the positive and negative directions corresponding with each axis (X, Y, and Z) of the space.

\subsection{Cache Lines}

When a program running on a CPU accesses memory, it can only retrieve data of a specific size. This is the cache line size, and in most modern computers it is 64 Bytes. If, for example, the program requires the value of a 4-Byte data structure, the CPU would retrieve the entire cache line where this value resides. After the CPU accesses the line it is stored in cache, a memory structure that exists physically closer to the CPU and, therefore, has a much shorter access time.

When data structures span the boundaries between cache lines, they require multiple memory accesses to fully read since the cache lines must be fetched one at a time from memory. For this reason, compilers prioritize aligning data structures to avoid bridging multiple lines where possible. While this ``wastes'' or fragments memory, this method is overall faster since it generally leads to significantly fewer memory accesses on average. By intentionally designing data structures to be an integer divisor of the cache line size, alignment can be easily actualized while also maintaining full use of the memory.

\section{Proposed Optimizations}

\subsection{Internal Connectivity}

The first proposed strategy is relatively simple: give each vertex its own 64-Byte cache line to store its Cartesian coordinates and neighbor data. Since the Cartesian coordinate data is almost always accessed alongside the neighbor data, it benefits the program to co-locate them in memory. That means both the coordinated data and neighbor values fit into a single data structure on one cache line. The coordinate values are each 8-Byte floating point numbers, for a total of 24 Bytes dedicated to spatial information. The other 40 Bytes can be dedicated to storing neighbor information. The neighbor indices are stored as 16-bit unsigned integers. This provides a tradeoff in the total number of possible vertices and the maximum number of neighbors. Using 16 bits for indexing thus limits the total number of vertices to a maximum of 65,536. In testing, there were no shapes that contained more vertices than this limit.

\begin{figure}
    \centering
    \includegraphics[width=1\linewidth]{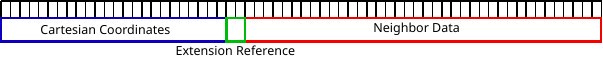}
    \caption{Memory Layout of the Vertex-Traversing Hull Data Structure. Each Cell is one Byte.}
    \label{fig:enter-label}
\end{figure}

To support vertices with more than twenty neighbors, an ``extension'' structure is implemented wherein each extension contains a potential reference to the next extension and memory space for fifteen more neighbor references. This allows support for relatively common shapes that would otherwise exceed this twenty-neighbor limit, such as a convex hull approximating a cone. The vertex must hold a 16-bit value referencing this extension, so the original structure can now only hold references to nineteen neighbors, with any more added to the extension.

Since not all vertices will have nineteen neighbors, what happens to the extra allocated space in the data structure? Instead of leaving it unallocated, the vertex can be allocated additional ``artificial'' edges that serve to connect it to vertices it does not actually share an edge with. These edges allow the hill climbing algorithm to ``skip'' intermediate vertices and take significantly fewer steps along the path to the support vertex. However, this raises another question: how are these artificial neighbors selected?

The candidate artificial neighbor vertices are ranked by their distance to the surface of a sphere centered around the target vertex, with the best vertices by this metric added to the neighbor list as space allows. The number added corresponds to the amount required to use up the rest of the space in the original array. if the vertex has less than nineteen real neighbors, artificial neighbors are added to pad out the original data structure. If the vertex has more than nineteen real neighbors, artificial neighbors are added to use the rest of the space in the last required extension array. In testing, a good value for the radius of this sphere was one-fifth of the radius of the overall bounding sphere for the convex hull. 

More complex heuristics than this bounding sphere, such as connectivity based on clustering algorithms, may provide better results. These may be evaluated in a future publication.

\begin{figure}
    \centering
    \includegraphics[width=0.7\linewidth]{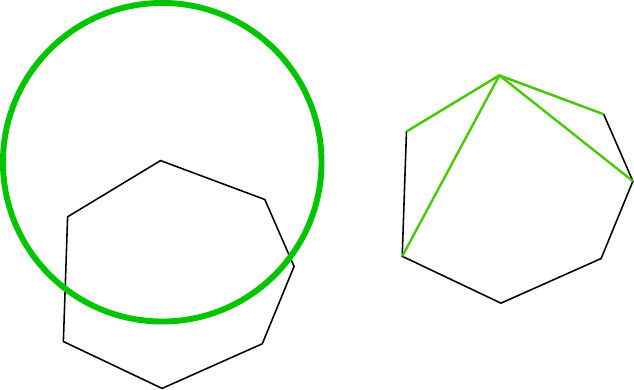}
    \caption{Interior Connections as Decided by Enclosing Sphere}
    \label{fig:enter-label}
\end{figure}

\subsection{Face Normal Traversal}

As discussed in the section on precomputation, it is often viable to calculate an acceptable starting vertex using a computationally inexpensive approximate approach, then switch to an exact vertex-centric method that uses the approximate starting vertex as its initial state. This saves time if the heuristic is sufficiently fast and accurate enough to outperform simply using the vertex-centric approach from the start. By using face normal alignment with a search vector as a proxy, the proposed approach traverses the faces of the convex hull to find the face normal that has maximal dot product with the search vector.

This approach may be considered as solving the support point function on a ``dual polytope'' wherein the face normals are represented as points on the surface of a sphere. These points are connected according to which faces share an edge. This heuristic becomes more accurate the more closely the original convex hull approximates a sphere.

One benefit of the face-traversing approach is the inherent fixed connectivity structure. For a convex hull with triangular faces, each face has exactly three neighbors. Unlike the vertex-based structure, there is no need to provide extra space for a large number of potential neighbors; instead, the data structure can be truncated to a smaller size. Namely: 24 Bytes for the normal, 6 Bytes for the indices of neighboring faces, and 2 Bytes for the index of one corresponding vertex, for a total of 32 Bytes. As a result, two of these data structures may be packed per cache line.

\begin{figure}
    \centering
    \includegraphics[width=1\linewidth]{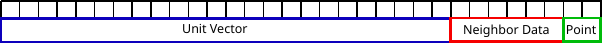}
    \caption{Memory Layout of Face-Traversing Hull Data Structure. Each Cell is one Byte.}
    \label{fig:enter-label}
\end{figure}

Since this is simply a heuristic, the algorithm still needs to transition to the vertex-traversing method after the optimal face is found. The data structure must store not only the face normal and connectivity data, but also the vertex coordinate and connectivity data. This leads to a significantly larger overall data footprint.

\subsection{Spherical Encoding}

The face normals are unit vectors, and can be represented as two-value spherical coordinates. This representation already enables a significant reduction in the size of the data structure. This can be taken further by observing that these values occur within a fixed, known range. The two spherical coordinate angles are stored as 4-Byte, Q1.31 fixed-point numbers \cite{yates_fixed-point_2001}. This means that the vector is encoded in only 8 bytes, rather than 24 Bytes. This brings the total size of the data structure down to 16 Bytes, meaning that four such data structures can fit into a single cache line.

\begin{figure}
    \centering
    \includegraphics[width=1\linewidth]{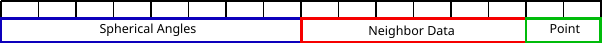}
    \caption{Memory Layout of Spherical-Encoded Face-Traversing Hull Data Structure. Each Cell is one Byte.}
    \label{fig:enter-label}
\end{figure}

Converting these values from angles back into vectors requires expensive trigonometry operations, the implementation details of which vary wildly between different computer architectures. To avoid this potential variance, a simple quadratic approximation of the Sine function that yields efficient results is implemented. The Cosine function is implemented by shifting the input value by pi/2 and evaluating the Sine function, and therefore has the exact same error profile as shown, just shifted by $\frac{\pi}{2}$. By exploiting the sign function, this yields a quadratic approximation of the Sine function that is accurate in the interval $[-\pi,\pi]$. The range of the Q1.31 value in fact spans this $[-\pi,\pi]$ interval. The equation is shown below in the range [-1, 1] for simplicity. 

\[sin(x \cdot \pi) \approx 4(x - sign(x) \cdot x^2)\]

\begin{figure}
    \centering
    \includegraphics[width=0.8\linewidth]{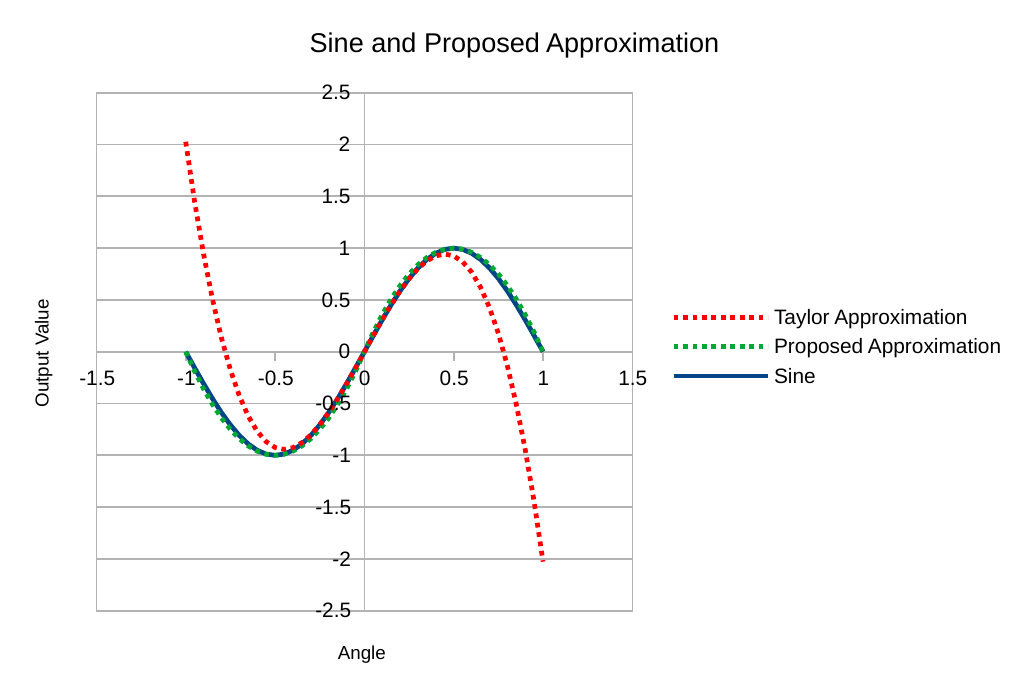}
    \caption{Plot of Sine and the Proposed Sine Approximation}
    \label{fig:SineApprox}
\end{figure}

\begin{figure}
    \centering
    \includegraphics[width=0.8\linewidth]{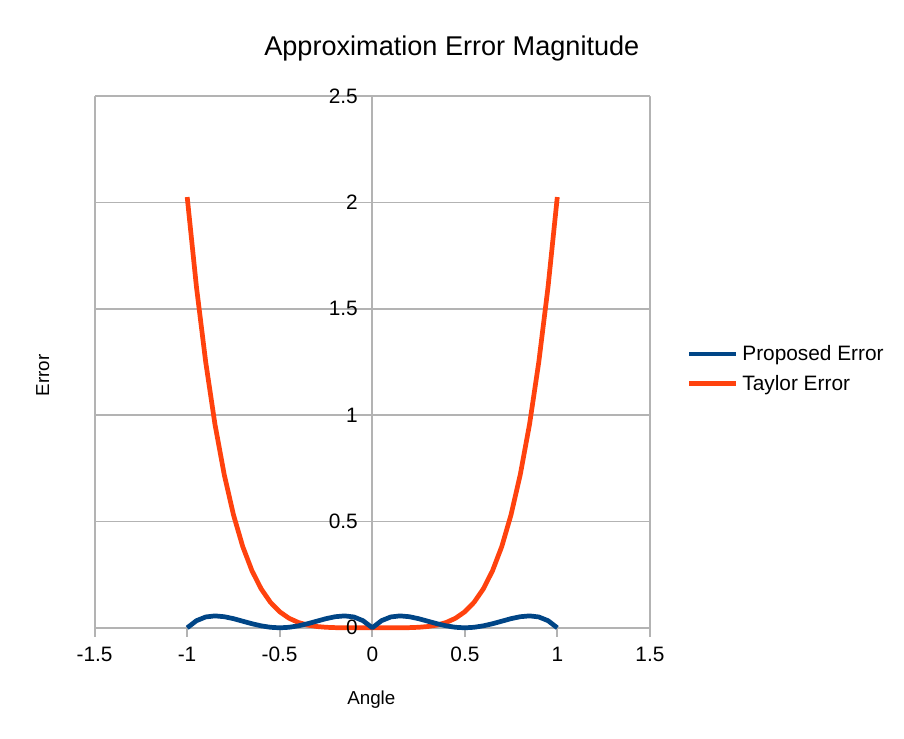}
    \caption{Approximation Error}
    \label{fig:ApproxError}
\end{figure}

One benefit of spanning the $[-\pi,\pi]$ interval with a fixed point data structure means that overflow and underflow results are still valid, since Sine and Cosine are periodic over this interval.

Another popular approximation for these trigonometric functions is to utilize the first few terms of their Taylor Series Expansion about zero. While this provides superior accuracy close to zero, this approximation leads to significantly higher inaccuracy at the outer limits of the function. The proposed approach, in contrast, produces reasonably low error across the entire domain of input values.

The proposed approximation and the Taylor Series approximation are plotted against the Sine function in Figure \ref{fig:SineApprox}, with the error in the proposed approximation shown in Figure \ref{fig:ApproxError}.

\section{Results}

\subsection{Methodology}

I implemented the proposed strategies in the Rust programming language. This language was selected for its consistent memory layout and lack of garbage collection; other languages, such as C or C++, could have been equivalently used, since they also satisfy these constraints. Additionally, the Rust compiler uses an LLVM backend, so other compilers with this backend, such as Clang, are expected to produce similar results. The code is licensed under the MIT License and is freely available at this repository: \url{https://sr.ht/~michaelgreer/Convex_Hull_Cache_Optimizations/}.

The methods in this paper were tested against a Rust-native collision framework called Parry, and a method matching the Bullet Hill Climbing process was implemented for a comparison with the existing state of the art.

The code was run on three different systems to illustrate the effect different cache sizes had on runtime. These systems were a Raspberry Pi Zero, a Framework 13 Laptop, and an HP Proliant Enterprise Server. Details of the systems and their cache sizes may be found in the table below.

\begin{footnotesize}
\begin{center}
\begin{tabular}{lccccl}

\toprule

\textbf{System} & \textbf{Processor} & \textbf{Clock Speed} & \textbf{L1 Cache} \\

\midrule

Pi Zero & BCM2835 & 0.7 GHz & 16KB \\

Framework & i5-1135G7 & 4.2 GHz & 192KB \\

HP Proliant & Xeon E5-2680 v3 & 3.3 GHz & 768KB \\

\bottomrule

\end{tabular}
\end{center}
\end{footnotesize}
For basic testing, convex hulls were generated by sampling points on a sphere and generating the convex hull using the Rust Parry Library. Using this generation method, the number of points on the resultant convex hull could be precisely controlled. For a more real-world application, several meshes from the Stanford 3D Scan Repository (\url{https://graphics.stanford.edu/data/3Dscanrep/}) were also included and tested.

Results from three applications are presented: standalone support point queries, integrated GJK algorithm tests, and, lastly, a robotic path planning example.

\subsection{Support Point Queries}

The proposed changes perform well compared to the two baselines, Parry and Bullet, with the internally connected method performing the most optimally. Interestingly, the trend lines for the Intel i5 CPU performance suggest that, for excessively large convex hulls, calculating the dot product on a per-vertex basis may perform better than hill climbing. 

\begin{figure}
    \centering
    \includegraphics[width=0.7\linewidth]{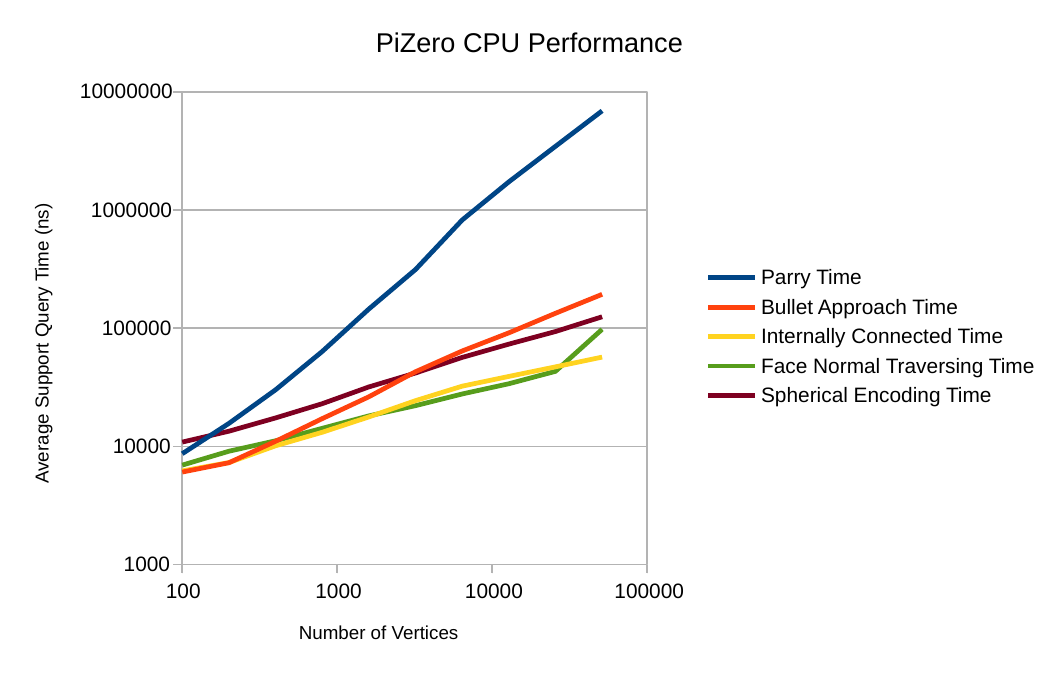}
    \caption{Timing Results for Raspberry Pi Zero}
    \label{fig:enter-label}
\end{figure}

\begin{figure}
    \centering
    \includegraphics[width=0.7\linewidth]{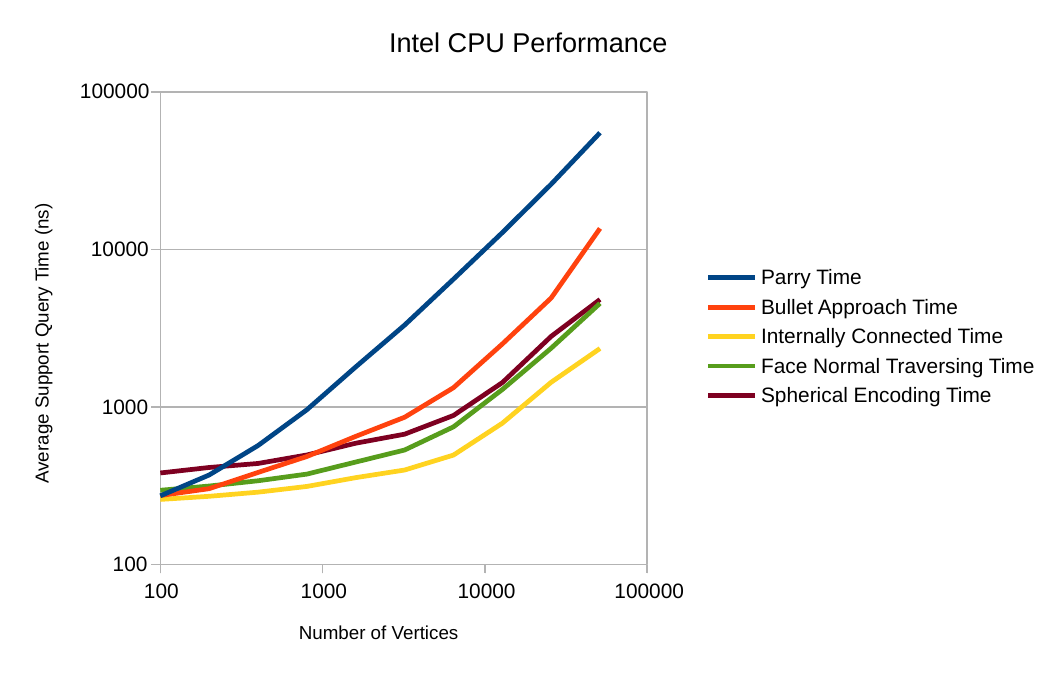}
    \caption{Timing Results for Framework}
    \label{fig:enter-label}
\end{figure}

\begin{figure}
    \centering
    \includegraphics[width=0.7\linewidth]{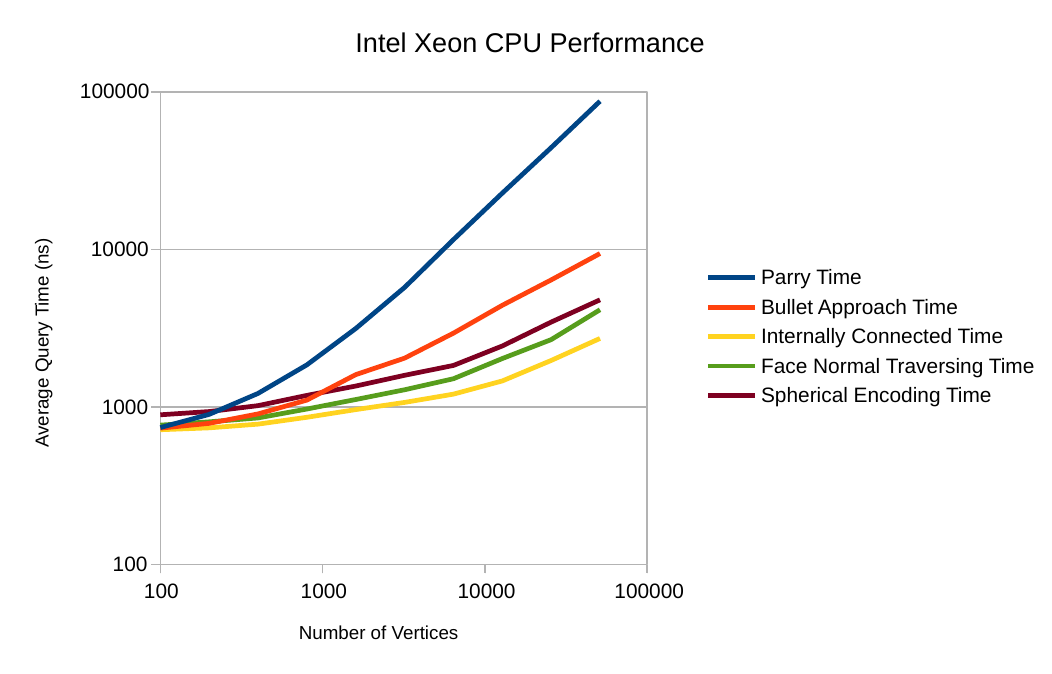}
    \caption{Timing Results for HP Proliant}
    \label{fig:enter-label}
\end{figure}

\begin{center}
\begin{tabular}{lccccl}

\toprule

\textbf{Method} & \textbf{Armadillo} & \textbf{Dragon} & \textbf{Lucy} \\

\midrule

Parry & 2171ns & 2767ns & 1881ns \\

Bullet & 531ns & 633ns & 587ns \\

Internally Connected & 376ns & 460ns & 487ns \\

Face Traversing & 443ns & 482ns & 513ns \\

Spherical Encoded & 567ns & 644ns & 723ns \\

\bottomrule

\end{tabular}
\end{center}

Support queries are frequently used in tandem with the GJK algorithm, but they are also often used on their own in the calculation of other collision structures like axis-aligned bounding boxes (AABBs).

\subsection{GJK Proximity Queries}

These support point strategies must perform well not only on their own, but also in the context of the GJK algorithm. An implementation of the Signed Volumes GJK algorithm \cite{montanari_improving_nodate} was used to benchmark how the support point functions perform for shapes that are in collision, close together, and distant from one another.

\begin{figure}
    \centering
    \includegraphics[width=0.7\linewidth]{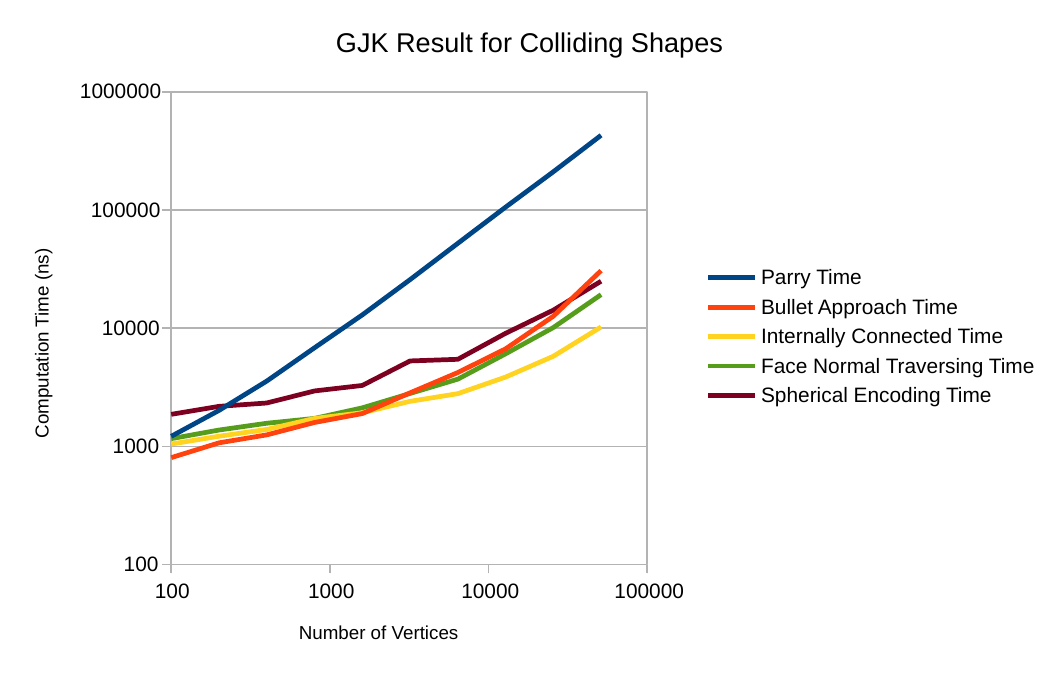}
    \caption{GJK Timing Results for Colliding Shapes}
    \label{fig:enter-label}
\end{figure}

\begin{figure}
    \centering
    \includegraphics[width=0.7\linewidth]{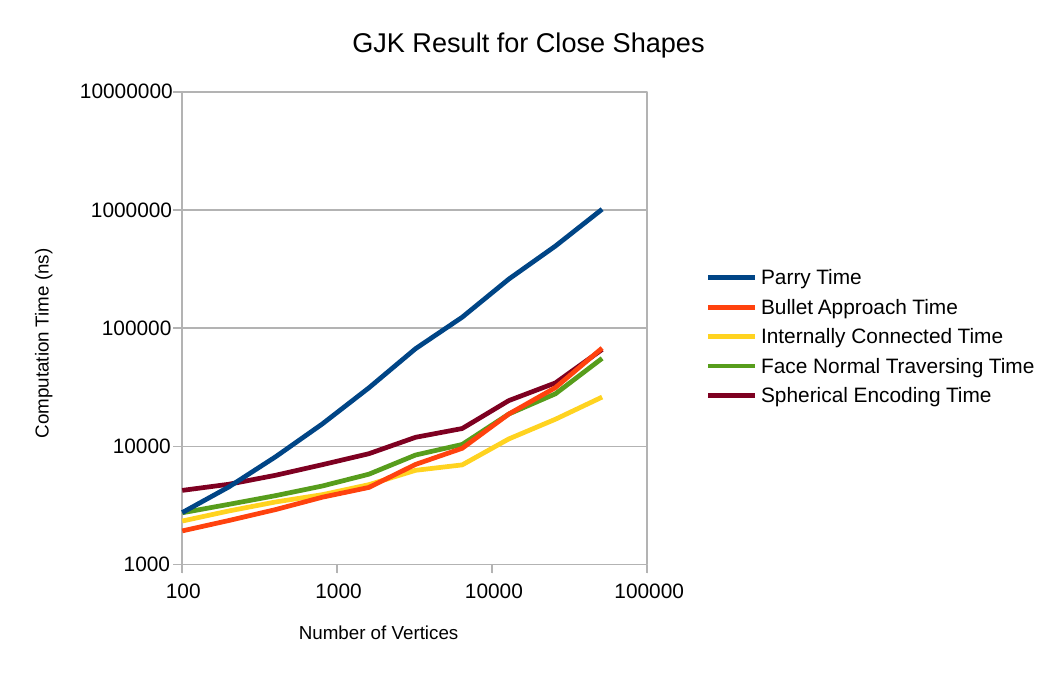}
    \caption{GJK Timing Results for Close Shapes}
    \label{fig:enter-label}
\end{figure}

\begin{figure}
    \centering
    \includegraphics[width=0.7\linewidth]{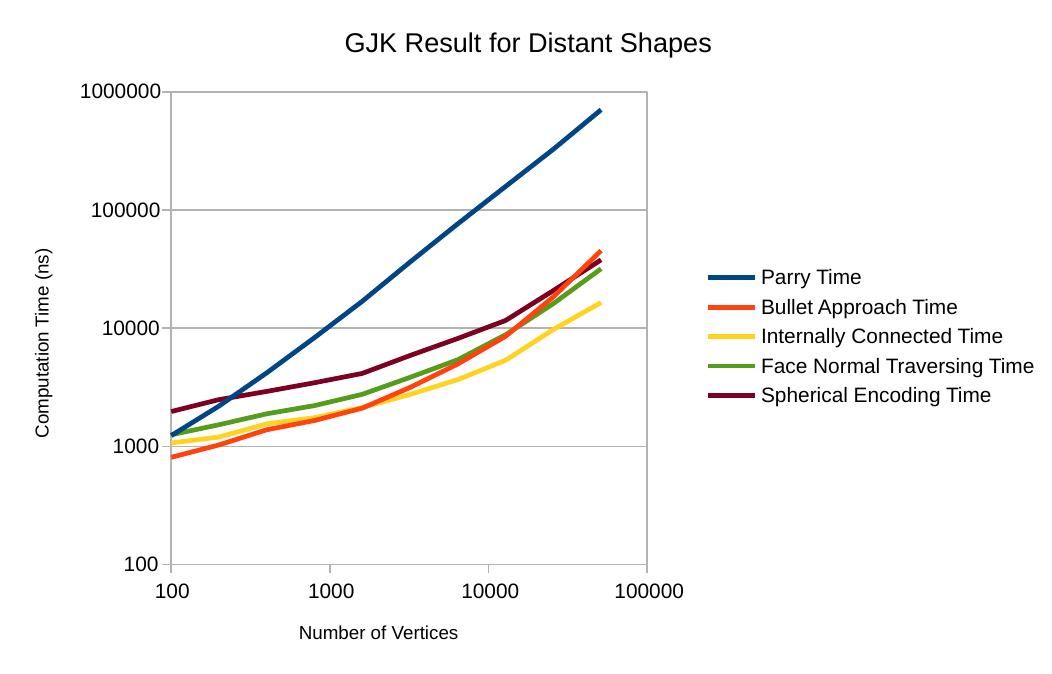}
    \caption{GJK Timing Results for Distant Shapes}
    \label{fig:enter-label}
\end{figure}

The Bullet-style approach outperforms all of the proposed methods for small (less than 2000 vertices) convex hulls, but the internally connected method performs better for more detailed convex hulls. This is likely due to the smaller overall memory footprint of the Bullet approach dominating for smaller shapes, while the interior connections allow the points to skip a significant number of vertices for larger shapes. While the existing Bullet approach is advantageous for smaller hulls, the proposed approach is significantly faster for more detailed convex hulls. This suggests an adaptive approach that dynamically selects the appropriate data structure based on the number of vertices may be most successful. For example, an algorithm that uses a traditional support point approach for hulls containing fewer than 1000 vertices, but switches to the internally connected approach for hulls containing larger than 1000 vertices.

\begin{figure*}
    \centering
    \includegraphics[width=0.8\linewidth]{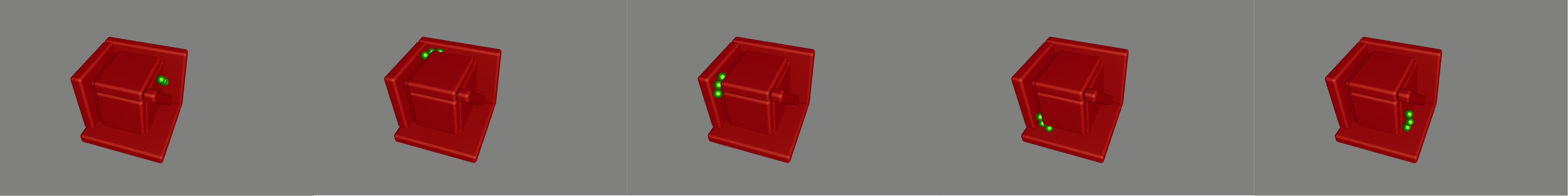}
    \caption{A Snake-Like Robot Navigating a Series of Tight Corridors}
    \label{fig:snake-robot}
\end{figure*}

\subsection{RRT Path Planning}

To demonstrate the applications of this approach in a robotics context, I implemented a basic path planning example of a snake-like robot navigating an environment consisting of several narrow passages, shown in Figure \ref{fig:snake-robot}. The models that enclose the corridors are omitted for viewing purposes, but are present in the collision representation of the scene. The RRT algorithm \cite{1573950399665672960} is used to perform path planning, while collisions are checked using the methods shown above. Since, in the earlier GJK results, only the internally connected approach was competitive with the Bullet approach, only the internally connected approach and the Bullet approach were tested using the RRT algorithm. 

The RRT algorithm was run ten times to reduce variances, with the fastest, average, and slowest times shown. The times include only the time to perform collision checking; the time consumed by the RRT algorithm is not included. The times displayed shown are the time it takes to perform a single GJK iteration.

\begin{center}
\begin{tabular}{lccccl}

\toprule

\textbf{Method} & \textbf{Fastest} & \textbf{Average} & \textbf{Slowest} \\

\midrule

Bullet & 8748ns & 10680ms & 12447ns \\

Internally Connected & 7766ns & 8518ns & 10431ns \\

\bottomrule

\end{tabular}
\end{center}

On average, the proposed internally connected approach achieves a 20.27\% improvement in computational time compared to the Bullet approach. Both the green spheres and the red rounded rectangular prisms have in the order of $10^5$ vertices, therefore, this result is consistent with the support point and GJK computation times shown earlier. A video of the RRT path solution is available here: \href{https://youtu.be/NChE60MNr88}{https://youtu.be/NChE60MNr88}.

\section{Conclusion}

In many conditions, the internally connected vertex-walking strategy provides significant improvements along the entire range of convex hull sizes, when compared to both existing approaches and the other proposed strategies. In highly constrained environments, the face-walking approach performs marginally better for certain convex hull sizes, but, overall, it is unclear if this difference in performance is a worthwhile topic of exploration.

The proposed methods have the benefit of being fully abstracted behind the support-point interface used by GJK implementations, so minimal refactoring work is necessary to integrate these optimizations into existing libraries. 

Future work will explore additional ways to use widely available CPU features to improve performance, such as SIMD operations and (increasingly standard) tensor processing units.

\section{Acknowledgments}

I would like to thank the reviewers of this paper for their insightful feedback on ways to improve this work. I would also like to thank Jordan Aronson for his generous gift of the hardware used to run test cases.

\bibliographystyle{plain} 
\bibliography{root}

\end{document}